\documentclass[12pt,letterpaper]{revtex4}
\usepackage{wrapfig}
\usepackage{graphicx,setspace,color,endnotes}
\usepackage[format=plain, font={footnotesize, stretch=0.9}, justification=centerlast, labelfont={footnotesize, bf}]{caption}
\usepackage{txfonts}

\let\footnote=\endnote

\newcommand{\pot}[2]{#1 \times 10^{#2}}

\begin{document}

\begin{center}

{\bf\LARGE Astro2020 Science White Paper}

\vspace{2mm}

{\bf\Large Spectral Distortions of the CMB as a Probe of Inflation, Recombination, Structure Formation and Particle Physics}

\end{center}

\vspace{-2mm}

\hspace{17mm} {\bf Primary thematic area:} Cosmology and Fundamental Physics

\hspace{17mm} {\bf Secondary thematic area:} Galaxy Evolution

\hspace{17mm} 
{\bf Corresponding author email:} Jens.Chluba@Manchester.ac.uk

\vspace{-2mm}

\thispagestyle{empty}

\begin{center}
{\small
J.~Chluba$^{1}$,
A.~Kogut$^{2}$,
S.~P.~Patil$^{3}$,
M.~H.~Abitbol$^{4}$,
N.~Aghanim$^{5}$,
Y.~Ali-Ha\"imoud$^{6}$,
M.~A.~Amin$^{7}$,
J.~Aumont$^{8}$,
N.~Bartolo$^{9,10,11}$,
K.~Basu$^{12}$
% ...
}\\[0mm]

\vspace{3mm}

{\bf \large Additional endorsers:}  \url{www.Chluba.de/SDWP-Decadal-2020/Endorsers.pdf}

\vspace{1mm}

\end{center}

\vspace{-4mm}

\noindent
{\scriptsize
$^1$ Jodrell Bank Centre for Astrophysics, School of Physics and Astronomy, The University of Manchester, Manchester M13 9PL, U.K.
\\[-2mm]
$^2$ NASA/GSFC, Mail Code: 665, Greenbelt, MD 20771, USA
\\[-2mm]
$^3$ Niels Bohr International Academy and Discovery Center, Blegdamsvej 17, 2100 Copenhagen, Denmark
\\[-2mm]
%
$^4$ University of Oxford, Denys Wilkinson Building, Keble Road, Oxford, OX1 3RH, UK
\\[-2mm]
%
$^5$ Institut d'Astrophysique Spatiale (IAS), CNRS (UMR8617), Universit\'e Paris-Sud, Batiment 121, 91405 Orsay, France
\\[-2mm]
%
$^6$ Center for Cosmology and Particle Physics, Department of Physics, New York University, New York, NY 10003, USA
\\[-2mm]
%
$^7$ Department of Physics and Astronomy, Rice University, 6100 Main Street, Houston, Texas 77005, USA
\\[-2mm]
%
$^8$ IRAP, Universite de Toulouse, CNRS, CNES, UPS, France
\\[-2mm]
%
$^9$ Dipartimento di Fisica e Astronomia 'Galileo Galilei', Universita' degli Studi di Padova, via Marzolo 8, I-35131, Padova, Italy
\\[-2mm]
%
$^{10}$ INFN, Sezione di Padova, via Marzolo 8, I-35131, Padova, Italy
\\[-2mm]
%
$^{11}$ INAF-Osservatorio Astronomico di Padova, Vicolo dell'Osservatorio 5, I-35122 Padova, Italy
\\[-2mm]
$^{12}$ Argelander-Institut f\"ur Astronomie, Universit\"at Bonn, Auf dem H\"ugel 71, D-53121 Bonn, Germany
\\[-2mm]
%
% ...
}
\newpage
\setcounter{page}{0}
\thispagestyle{empty}

\section*{EXECUTIVE SUMMARY}
\vspace{-4.5mm}
Following the pioneering observations with COBE in the early 1990s, studies of the cosmic microwave background (CMB) have focused on temperature and polarization anisotropies.
CMB spectral distortions -- tiny departures of the CMB energy spectrum from that of a perfect blackbody -- provide a second, independent probe of fundamental physics, with a reach deep into the primordial Universe. The theoretical foundation of spectral distortions has seen major advances in recent years, which highlight the immense potential of this emerging field. Spectral distortions probe a fundamental property of the Universe -- its {\it thermal history} -- thereby providing additional insight into processes within the cosmological standard model\footnote{When referring to the cosmological standard model (CSM) we assume the $\Lambda$CDM parameterization, supplemented by the Standard Model of particle physics, admitting that the presence of dark matter and dark energy requires physics beyond the latter.} (CSM) as well as new physics beyond.
Spectral distortions are an important tool for understanding inflation and the nature of dark matter. They shed new light on the physics of recombination and reionization, both prominent stages in the evolution of our Universe, and furnish critical information on baryonic feedback processes, in addition to probing primordial correlation functions at scales inaccessible to other tracers. In principle the range of signals is vast: {\it many orders of magnitude of discovery space} could be explored by detailed observations of the CMB energy spectrum. 
Several CSM signals are predicted and provide clear experimental targets, some of which are already observable with present-day technology. 
Confirmation of these signals would extend the reach of the CSM by orders of magnitude in physical scale as the Universe evolves from the initial stages to its present form. {\it The absence of these signals would pose a huge theoretical challenge, immediately pointing to new physics.} 

\vspace{-5mm}
\section{Cosmology beyond thermal equilibrium}
\vspace{-4.5mm}
Cosmology is now a precise scientific discipline, with detailed theoretical models that fit a wealth of very accurate measurements. Of the many cosmological data sets, the CMB temperature and polarization anisotropies provide the most stringent and robust constraints on
theoretical models, allowing us to determine the key parameters of our Universe (e.g., the total density, expansion rate and baryon content) with unprecedented precision, while simultaneously addressing fundamental questions about inflation and early-universe physics. By studying the statistics of the CMB anisotropies with different experiments over the past decades we have entered the era of precision cosmology, clearly establishing the highly-successful $\Lambda$CDM concordance model \cite{Smoot1992, WMAP_params, Planck2013params}.

But the quest continues. Despite its many successes, $\Lambda$CDM {\it is known to be incomplete}. It traces the growth of structure in the Universe from primordial density perturbations to the modern era, but the origin of those perturbations remains poorly understood. In addition, in spite of relentless efforts, the nature of dark matter (DM) and dark energy remains a mystery. Together, these enigmatic components comprise 95\% of the energy density of the Universe. Particle and high-energy physics offer candidate solutions of these problems (e.g., inflation and particle dark matter), but these inevitably require new physics beyond the Standard Model of particle physics.

{\it Precision measurements of the} CMB {\it energy spectrum open a new window into the physics of the early Universe, constraining models in ways not possible using other techniques.} Departures of the CMB energy spectrum from a pure blackbody -- commonly referred to as {\it spectral distortions} -- encode unique information about the thermal history of the Universe, from when it was a few months old until today. Since the measurements with COBE/FIRAS in the early '90s, the sky-averaged CMB spectrum is known to be extremely close to a perfect blackbody at a temperature $T_0=(2.7255\pm 0.0006)\,{\rm K}$ \cite{Fixsen1996, Fixsen2009}, with possible distortions limited to one part in $10^5$. This impressive measurement was awarded the 2006 Nobel Prize in Physics and already rules out cosmologies with extended periods of large energy release. 

\begin{wrapfigure}[20]{l}{0.62\textwidth}
\vspace{-2.5mm}
   \includegraphics[width=0.62\columnwidth]{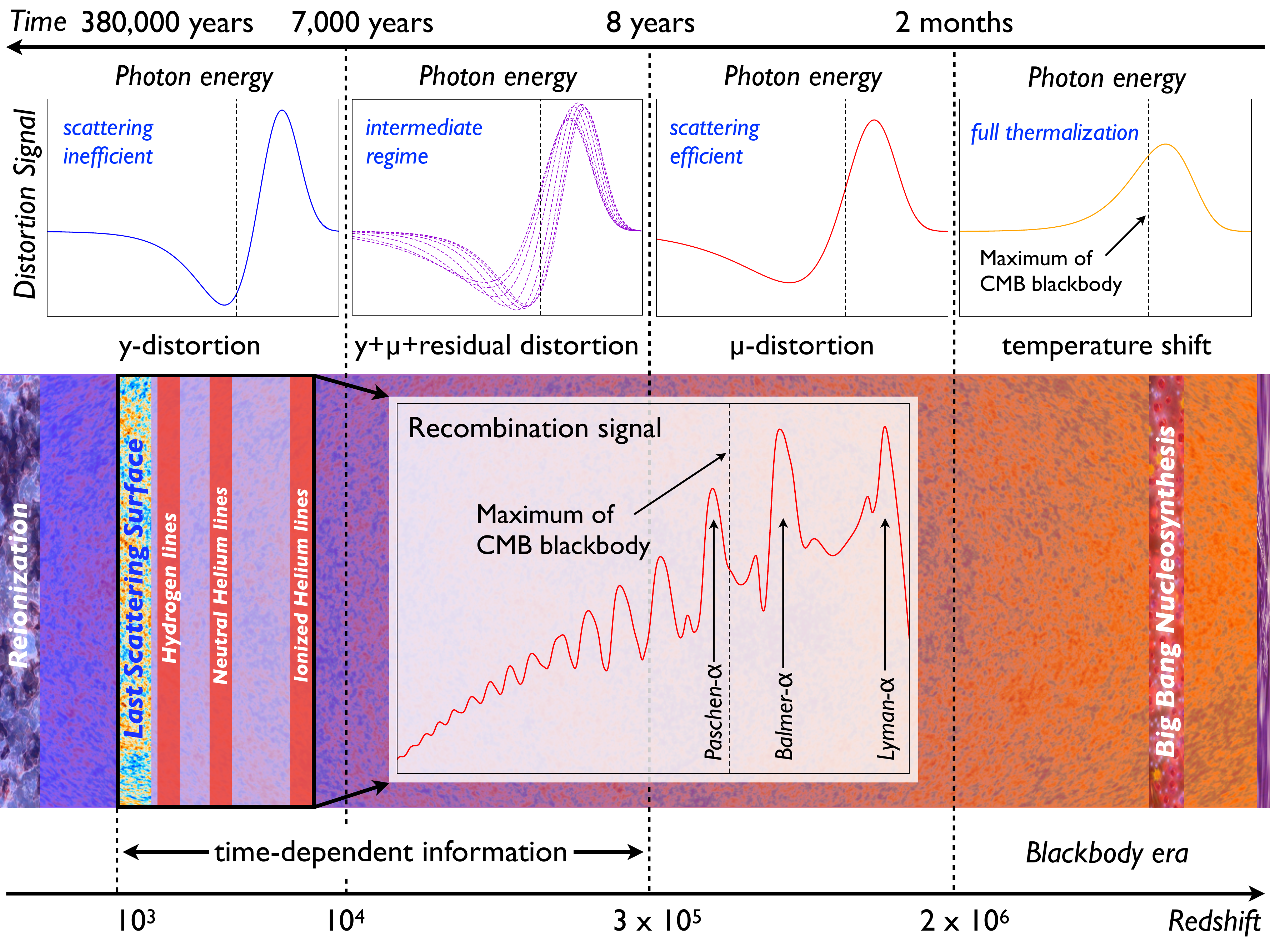}
   \caption{Evolution of spectral distortions across time. Distortions probe the thermal history over long periods deep into the primordial Universe that are inaccessible by other means.}
   \label{fig:stages}
\end{wrapfigure}

Spectral distortions are created by processes that drive matter and radiation out of thermal equilibrium after thermalization becomes inefficient at redshift $z\lesssim \pot{2}{6}$. Examples are energy-releasing mechanisms that heat the baryonic matter, inject photons or other electromagnetically-interacting particles. The associated signals are usually characterized as $\mu$- and $y$-type distortions, formed by energy exchange between electrons and photons through Compton scattering \citep{Zeldovich1969, Sunyaev1970mu, Illarionov1975b, Burigana1991, Hu1993}. Compton scattering is inefficient at $z \lesssim \pot{5}{4}$, yielding a $y$-type distortion, which probes the thermal history during recombination and reionization (Fig.~\ref{fig:stages}). 
In contrast, a $\mu$-type (or chemical potential) distortion forms at $z\gtrsim \pot{5}{4}$, when Compton scattering is very efficient. A $\mu$-distortion cannot be generated at recent epochs and thus directly probes events in the pre-recombination era. 

The simple classical picture has been refined in recent years. We now understand that the transition between $\mu$- and $y$-type distortions is gradual and that the signal contains additional time-dependent information \citep{Chluba2011therm, Khatri2012mix, Chluba2013Green}. This extra information is contained in the residual or $r$-type distortion and can be used to distinguish energy release mechanisms \citep{Chluba2013fore, Chluba2013PCA}. Distortions created by photon-injection mechanisms exhibit rich spectral phenomenology \citep{Chluba2015GreensII}, one prominent example being the distortion created by the cosmological recombination process \citep{Dubrovich1975, Sunyaev2009, Chluba2016CosmoSpec} (see Fig.~\ref{fig:stages}). Additional information can be imprinted by non-equilibrium processes in the pre-recombination hydrogen and helium plasma \citep{Liubarskii83, Chluba2008c, Chluba2010a} or by non-thermal particles in high-energy particle cascades \citep[e.g.,][]{Ensslin2000, Chluba2010a, Chluba2015GreensII, Slatyer2015, Acharya2018}. Finally, distortion anisotropies are expected and can be correlated against tracers of both primordial density perturbations and large-scale structure to further probe cosmic evolution \citep{Refregier2000, Zhang2004, Pitrou2010, Pajer2012, Ganc2012, Hill2013}.

Spectral distortions thus provide more than just a simple integral constraint for cosmology. They are a unique and powerful probe of a wide range of interactions between particles and CMB photons, reaching back all the way from the present to a few months after the Big Bang and allowing us to access information that cannot be extracted in any other way. Broad overviews of the CMB spectral distortion science case can be found in \citep{Sunyaev2009, Chluba2011therm, Sunyaev2013, PRISM2013WPII, Chluba2014Moriond, Tashiro2014, deZotti2015, Chluba2016, Chluba2018}.

\vspace{-6mm}
\section{Physics Beyond $\Lambda$CDM and the Standard Model}
\vspace{-4.5mm}
A central question in modern cosmology is the origin of the observed primordial density perturbations. Measurements from CMB anisotropies and large-scale structure find a nearly scale-invariant power spectrum $\mathcal{P}(k) \simeq k^{n_{\rm S} - 1}$ with spectral index $n_{\rm S}= 0.965 \pm 0.004$, sampled over a range of spatial scales $k\simeq 10^{-4}$ to $0.1$ Mpc$^{-1}$ \citep{Planck2018params}. Their phase coherence is evidence for their super-Hubble nature, and their near scale-invariance is evidence of a weakly broken shift symmetry in the underlying theory. However, their precise origin is as of yet unknown. Inflation provides a widely accepted framework \citep{Starobinsky:1980te, Guth:1980zm, Linde:1981mu, Albrecht:1982wi}, with the simplest models generically predicting a small departure from scale-invariance (with $n_{\rm S} < 1$) as the inflaton rolls down its potential \citep{mukhanov, Hawking:1982cz, Starobinsky:1979ty, Guth:1982ec}.
However, various alternatives to inflation have been proposed \citep{Gasperini:1992em,Wands:1998yp, Khoury:2001bz, Hollands:2002yb, Brandenberger:2006vv, Craps:2007ch, Creminelli:2010ba, Easson2011, Ijjas:2016tpn, Dobre2018} and no clear theoretical consensus has yet emerged. Searches for a $B$-mode pattern in CMB polarization could yield compelling evidence for the simplest inflationary models. CMB polarization measurements so far only provide upper limits \cite{Aghanim:2018eyx, Ade:2018gkx} with no firm target from theory for a guaranteed detection. However, detection of a tensor to scalar ratio of $r \simeq 10^{-3}$ is a distinguishing benchmark for large-field models, which in certain realizations further manifest the specific relation $r \simeq (1 - n_{\rm S})^2$ \citep[e.g.,][]{Starobinsky:1980te, Bezrukov2008}.

\begin{wrapfigure}[18]{l}{0.62\textwidth}
\vspace{-4.0mm}
\hspace{-1mm}
\includegraphics[width=0.62\textwidth]{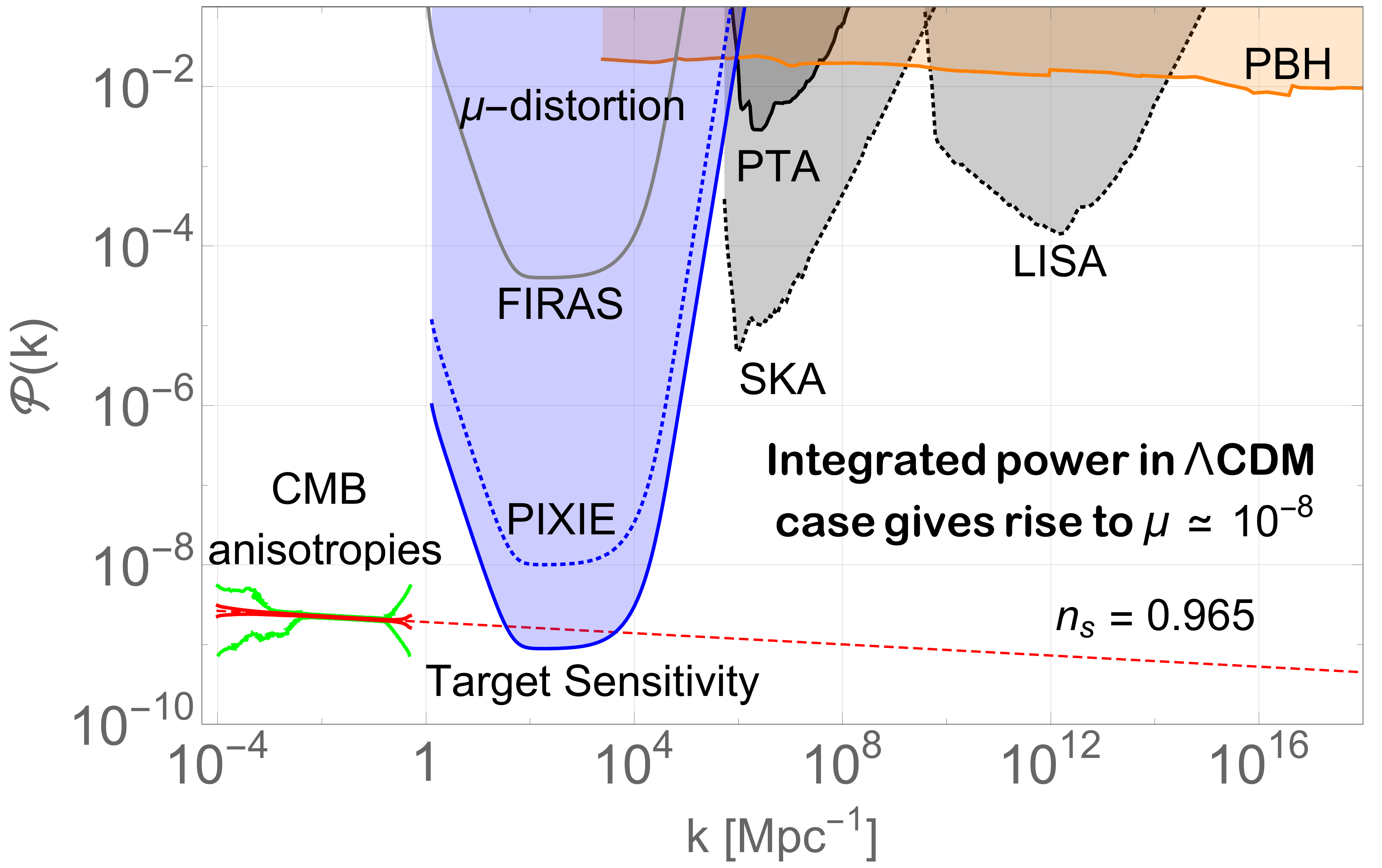}
\caption{Forecast constraints (95 \% c.l.) on the primordial power spectrum for features with a $k^4$ profile that cuts off sharply at some\protect\footnote{To avoid the unrealistic GW spectrum generated by a $\delta$-function scalar power spectrum, we plot all integrated constraints using a $k^4$ spectrum -- see \citep{Byrnes:2018txb} for the reason for this choice. The peak sensitivity for $\mu$-distortions is effectively unchanged were we to instead plot constraints for $\delta$-function features in the power spectrum with the same integrated power (see Fig.~9 therein), and also \cite{Inomata:2018epa}.} $k_p$ \citep[see][for more details]{Byrnes:2018txb}. $\mu$-distortions constrain perturbations at scales and levels inaccessible to other probes.}
\label{fig:forecast}
\end{wrapfigure}

{\it Spectral distortions provide a unique new probe of primordial density perturbations.}
Inflation may or may not be a valid description of the early Universe, but density perturbations are known to exist; regardless of their origin, dissipation of these perturbations through photon diffusion (Silk damping) in the early Universe 
will distort the CMB spectrum at observable levels \citep{Sunyaev1970diss, Daly1991, Hu1994, Chluba2012, Khatri2012short2x2}.
The signal can be accurately calculated using simple linear physics and depends on the amplitude of primordial perturbations at scales with $k\simeq 1-10^4\,{\rm Mpc^{-1}}$, some ten e-folds further than what can be probed by CMB anisotropies (Fig.~\ref{fig:forecast}). If the near scale-invariance of the power spectrum observed on large scales persists to these much smaller scales, then the predicted distortion, $\mu\simeq \pot{(2.3\pm0.14)}{-8}$ \citep{Chluba2012, Cabass2016, Chluba2016}, could be observed using current technology ($\S$IV). Detecting this signal extends our grasp on primordial density 
%
%
perturbations by over three orders of magnitude in scale, covering epochs that cannot be probed directly in any other way. 
{\it A non-detection at this level would be a serious challenge for} $\Lambda$CDM, {\it immediately requiring new physics.} 

Measurements of $\mu$-distortions are directly sensitive to the power spectrum amplitude and its scale-dependence around $k\simeq 10^3\,{\rm Mpc^{-1}}$ \citep{Chluba2012inflaton, Pajer2012b, Khatri2013forecast, Chluba2013PCA}. Within the slow-roll paradigm, this provides a handle on higher-order slow-roll parameters (i.e., running of the tilt), benefiting from a vastly extended lever arm  \citep{Powell2012, Clesse2014, Cabass2016SRparams}.
Outside of standard slow-roll inflation, large departures from scale-invariance are well-motivated and often produce excess small-scale power (e.g., features \citep{Starobinsky:1992ts, Adams:2001vc, Hazra:2014jka, Hazra:2017joc} or inflection points \citep{Polnarev:2006aa, Kohri:2007qn, Ido2010, Choudhury:2013woa, Clesse:2015wea, Germani:2017bcs} in the potential, particle production \citep{Barnaby:2009dd, Cook:2011hg, Battefeld:2013bfl, Dimastrogiovanni2017, Domcke:2018eki}, waterfall transitions \citep{Linde:1993cn, Lyth:1996kt,Juan1996PhRvD, Abolhasani:2010kn, Clesse:2014pna}, axion inflation~\citep{Barnaby2011, Barnaby2012, Meerburg:2012id}, etc.~\citep{Chluba2015IJMPD}), implying the presence of new physical scales that can be probed with spectral distortions (Fig.~\ref{fig:forecast}). 
In this respect, spectral distortions could establish a link to a possible primordial origin of the small-scale structure crisis \citep{Nakama2017, Cho2017}. 
They could also place limits on primordial black holes \citep{Carr2010, Pani2013, Nakama2017xvq}, testing their role in super-massive black-hole formation \citep{Kohri2014SMBH} and as DM candidate \citep{Juan1996PhRvD, Clesse2015PBH}, the latter seeing renewed interest after the first LIGO/Virgo merger events \citep{Bird2016, Clesse2017PBHs, Kohri2018}.
Spectral distortions are also created by the dissipation of small scale {\it tensor} perturbations \citep{Ota2014, Chluba2015} and depend on the perturbation-type (i.e., adiabatic vs. iso-curvature) \citep{Hu1994isocurv, Dent2012, Chluba2013iso, Haga2018}, providing ways to test inflation scenarios in uncharted territory. 
Spectral distortion anisotropies furthermore probe local-type primordial non-Gaussianity at small scales \citep{Pajer2012, Ganc2012, Biagetti2013, Ota2015aniso_iso, Razi2015, Khatri2015aniso, Chluba2017, Ota2016muE, Ravenni2017, Cabass:2018jgj}, an exciting direction that complements other probes and could shed light on multi-field inflation scenarios \citep{Dimastrogiovanni2016}.

Dark matter is another example of how spectral distortions probe new physics. Non-baryonic matter constitutes $\simeq25$\% of the energy density of the Universe, but its nature remains unknown. The long-favored WIMP-scenario is under increasing pressure \citep{Ahmed:2009zw, Aprile:2012nq, Angloher:2015ewa, Agnese:2015nto, Tan:2016zwf, Akerib:2016vxi}, and emphasis is gradually shifting focus towards alternatives, prominent examples being axions, sterile neutrinos, sub-GeV DM or primordial black holes \citep{Jungman1996, Feng2003PhRvL, Feng2003, Kusenko2009, Feng2010, Carr2010, Marsh2016Rev}.
To solve this puzzle, a multi-tracer approach that combines different particle physics and cosmological probes is needed. Measurements of the CMB anisotropies themselves have clearly helped to establish the presence of DM on cosmological scales and provided tight constraints on DM annihilation and decay \citep{Ellis1992,Adams1998nr, Chen2004, Padmanabhan2005, Galli2009, Slatyer2009, Slatyer2017, Poulin2017, Dienes2018} and DM-SM-interactions \citep{Wilkinson2014, Dvorkin2014, Wilkinson2014b, Gluscevic2018, Boddy2018, Boddy2018b}. However, for DM annihilation and decay CMB anisotropies quickly lose constraining power before recombination ($z\gtrsim 10^3$), being impeded by cosmic variance. 
Similarly, measurements of light-element abundances \citep{Ellis1992, Kawasaki2005, Jedamzik2008, Kawasaki2018}, which are only sensitive to non-thermal energy release above nuclear dissociation thresholds in the pre-recombination era \citep{Chluba2013PCA, Poulin2015Loop}, saturated their limits due to astrophysical uncertainties. {\it This is where CMB spectral distortions offer a valuable complementary probe.} 
For decaying particle scenarios, distortions are sensitive to particles with lifetimes $t\simeq 10^6-10^{12}\,{\rm s}$ 
\citep{Sarkar1984, Ellis1985, Kawasaki1986, Hu1993b, Chluba2011therm, Chluba2013PCA, Aalberts2018}, providing direct measurement of particle lifetimes via $r$-type distortions \citep{Chluba2013fore, Chluba2013PCA}. Similarly, annihilating particles can be constrained using distortions; $\mu$-distortions are sensitive to light particles ($m \lesssim 100$ keV) and complement $\gamma$-ray searches for heavier particles \citep{McDonald2001, Chluba2013fore}. The rich spectral information added by various non-thermal processes \citep{Liubarskii83, Chluba2008c, Chluba2010a, Chluba2015GreensII, Slatyer2015, Acharya2018} will allow us to glean even more information about the nature of dark matter. Significant theoretical work is required, although it is already clear that in addition to the aforementioned examples distortions can meaningfully probe scenarios involving axions \citep{Tashiro2013, Ejlli2013, Mukherjee2018}, gravitino decays \citep{Ellis1985, Dimastrogiovanni2015}, strings \citep{Ostriker1987, Tashiro2012b}, DM-SM-interactions \citep{Yacine2015DM, Diacoumis2017, Slatyer2017}, macros \citep{Kumar2018} and primordial magnetic fields \citep{Jedamzik2000, Sethi2005, Kunze2014, Wagstaff2015}. 

\vspace{-6mm}
\section{Precision Tests of $\Lambda$CDM and the Standard Model}
\vspace{-4.5mm}
Spectral distortions also enable new tests of the standard cosmological model. The cosmological recombination and reionization eras as well as the cosmic dark ages mark important transitions in the evolution of our Universe. The largest all-sky distortion signal is indeed caused by the reionization and structure-formation processes \citep{Sunyaev1972b, Hu1994pert, Cen1999, Refregier2000, Miniati2000, Oh2003}. Energy output from the first stars and shocks heats the baryons and electrons, which then up-scatter CMB photons to create a $y$-type distortion. The overall distortion is expected to reach $y\simeq \pot{{\rm few}}{-6}$ \citep{Refregier2000, Miniati2000, Khatri2015y, Hill2015}, only one order of magnitude below the current upper limit placed by COBE/FIRAS. Such a distortion {\it must} exist and provides a measurement of the total thermal energy in (ionized) baryons in the Universe.
A large part of the signal is due to halos with masses $M\simeq 10^{13}\,M_\odot$ containing gas with an electron temperature of $k T_{\rm e}\simeq 2\,{\rm keV}$. This causes a relativistic temperature correction \citep{Wright1979, Rephaeli1995, Sazonov1998, Itoh98, Challinor1998} to the $y$-distortion that can directly tell us about feedback mechanisms \citep{Hill2015}. The low-redshift $y$-distortion from reionization is furthermore anisotropic \citep[e.g.,][]{Refregier2000, Zhang2004, Pitrou2010} and thus opens new
opportunities for cross-correlation studies (e.g., with CMB and 21 cm tomography).

The cosmological recombination process causes another small but inevitable distortion of the CMB. Line emission from hydrogen and helium injects photons into the CMB, which after redshifting from $z\simeq 10^3$ is visible today as complex frequency structure in the microwave bands (Fig.~\ref{fig:stages}) \citep{Dubrovich1975, RybickiDell94, DubroVlad95, Kholu2005, Wong2006, Jose2006, Chluba2006, Jose2008, Yacine2013RecSpec}.
The cosmological recombination radiation (CRR) has a simple dependence on cosmological parameters and the dynamics of recombination; since it includes not only hydrogen but also two helium recombinations, it probes eras well beyond the last-scattering surface observed by CMB anisotropies \citep{Chluba2008T0, Sunyaev2009, Chluba2016CosmoSpec}.
Cosmological recombination is furthermore crucial for understanding and interpreting the CMB temperature and polarization anisotropies \citep{Zeldovich68, Peebles68, Sunyaev1970, Peebles1970}. It is thus critical to test our physical assumptions during this era \citep{Hu1995, Lewis2006, Jose2010, Shaw2011}. {\it The} CRR {\it provides one of the most direct methods to achieve this.}
The CRR also enables a pristine measurement of the primordial helium abundance, long before the first stars have formed. 
It would thus break cosmological parameter degeneracies, e.g., between the primordial helium abundance and neutrino number \citep{Chluba2016CosmoSpec}. Finally, interactions of CMB photons with atoms can imprint additional frequency-dependent signals 
through resonance \citep{Loeb2001, Zaldarriaga2002, Kaustuv2004, Jose2005, Carlos2005, Carlos2006, Carlos2007, Carlos2017} 
and Rayleigh scattering effects \citep{Yu2001, Lewis2013}, both of which provide independent ways for
learning about recombination, the dark ages and reionization.

\vspace{-5mm}
\section{The path forward with CMB spectral distortions}
\vspace{-4.5mm}
The seminal measurements of the CMB blackbody spectrum by COBE/FIRAS in the early '90s cemented the Hot Big Bang model by ruling out any energy release greater than $\Delta U / U \simeq 10^{-5}$ of the energy in CMB photons \citep{Mather1994, Fixsen1996, Fixsen2009}. Advances since then, in both detector technology and cryogenics, could improve this sensitivity by four orders of magnitude or more (e.g., with experimental concepts like PIXIE \citep{Kogut2011PIXIE, Kogut2016SPIE} or PRISM \citep{PRISM2013WPII}), opening an enormous discovery space for both predicted distortion signals and those caused by new physics.

COBE/FIRAS was not background limited; its sensitivity was set instead by phonon noise from the 1.4 K detector. Modern detectors, operating at $\simeq 0.1$ K, would have detector (dark) noise well 
\begin{wrapfigure}[20]{l}{0.6\textwidth}
\vspace{-4mm}
   \includegraphics[width=0.6\columnwidth]{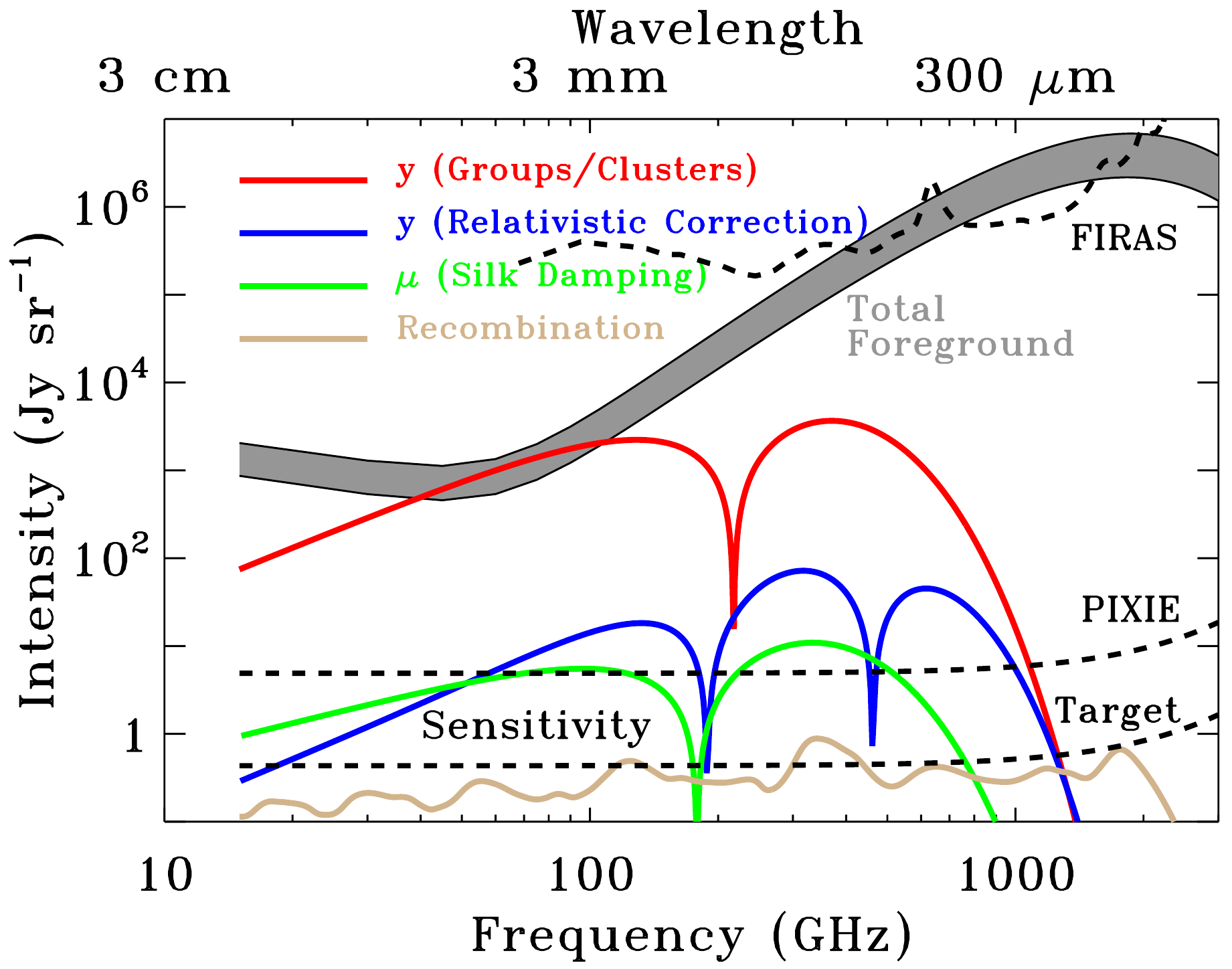}
   \caption{Spectral distortions are observable using current technology. The signal amplitudes and required foreground cleaning are comparable to planned $B$-mode searches.}
   \label{fig:future}
\end{wrapfigure}
below the intrinsic limit set by photon arrival statistics. The sensitivity of a background-limited instrument could be further improved by increasing the instrument's collecting area.

Figure \ref{fig:future} compares several predicted spectral distortions \citep[e.g.,][]{Chluba2016} and the largest astrophysical foregrounds\footnote{At high frequencies, these are dominated by dust from the galaxy and the cosmic infrared background, while at low frequencies it is synchrotron and free-free emission.}
to the sensitivity of possible next-generation spectrometers. 
Pioneering steps towards $y\simeq 10^{-7}$, yielding a clear detection of the expected average distortion caused by groups and clusters (see Sect.~III), are possible from the ground and  balloons (e.g., using concepts similar to COSMO, OLIMPO \citep{Masi2003, Schillaci2014} and ARCADE \citep{Kogut2006ARCADE, arcade2}). However, because the distortions peak at frequencies above 200 GHz, broad frequency coverage outside the atmospheric windows ultimately requires a space mission to detect $\mu\simeq 10^{-8}$ or the CRR \citep{Vince2015, Mayuri2015, Serra2016ApJ, Chluba2017Moments, abitbol_pixie, Remazeilles2018mu}. 
Both the anticipated signal levels (in the range 1--100 ${\rm Jy\,sr^{-1}}$) and level of foreground cleaning are comparable to those encountered for next-generation CMB polarization measurements \citep[e.g.,][]{Remazeilles2018}.
Therefore, much of the technology and analysis techniques are directly transferable, although a new synergistic approach (combining multiple data sets) and observing strategy (e.g., small-patch vs. all-sky) have yet to be developed.

To conclude, CMB spectral distortions probe many processes throughout the history of the Universe. Precision spectroscopy, possible with existing technology, would provide key tests for processes expected within the CSM and open an enormous discovery space for new physics. This offers unique scientific opportunities for furthering our understanding of inflation, recombination, reionization and particle physics. Many experimental and theoretical challenges have to be overcome before we can fully exploit this new window to early- and late-universe physics. However, the potential gains are immense and the field is entering a phase of accelerated growth after decades of dormancy. With a coordinated approach, the next decade could thus see new precision measurements of one of the fundamental observables of our Universe\footnote{We wish to acknowledge the generosity and support of the CERN Theoretical Physics Department in hosting the workshop "Probing fundamental physics with CMB spectral distortions" from Mar 12--16 2018 where this white paper was first conceived. This work has received funding from the European Research Council (ERC) under the European Union's
Horizon 2020 research and innovation program (grant agreement No 725456, CMBSPEC). JC is supported by the Royal Society as a Royal Society University Research Fellow at the University of Manchester, UK.}.

{\small
\vspace{-3mm}
\bibliographystyle{mn2e}
\bibliography{Lit}
}

\theendnotes

\end{document}